\documentclass[conference]{IEEEtran}

\AtBeginDocument{%
 \providecommand\BibTeX{{%
   \normalfont B\kern-0.5em{\scshape i\kern-0.25em b}\kern-0.8em\TeX}}}
    
\usepackage{xspace}

\usepackage{array}
\newcolumntype{L}{>{\arraybackslash}m{16cm}}
\newcolumntype{C}[1]{>{\centering\let\newline\\arraybackslash\hspace{0pt}}m{#1}}
\newcolumntype{R}[1]{>{\raggedleft\let\newline\\arraybackslash\hspace{0pt}}m{#1}}
\usepackage[numbered]{bookmark}
\usepackage{tcolorbox}
\usepackage{graphicx}
\usepackage{xcolor}
\usepackage{stfloats}
\usepackage[justification=centering]{caption}
\usepackage{lineno}
\usepackage{dirtytalk}
\usepackage{amsmath}
\usepackage{pgfplots, pgfplotstable}
\usetikzlibrary{patterns}

\usepackage[english]{babel}
\usepackage{textcomp}
\usepackage{mathtools}
\usepackage{ltxtable}
\usepackage{algorithm}
\usepackage{algpseudocode}
\usepackage{textcomp}

\usepackage{tcolorbox}

\usepackage{pgf-pie}
\definecolor{wedge1}{RGB}{ 190  30  46}
\definecolor{wedge2}{RGB}{ 240  65  54}
\definecolor{wedge3}{RGB}{ 241  90  43}
\definecolor{wedge4}{RGB}{ 247 148  30}
\definecolor{wedge5}{RGB}{  43  56 144}
\definecolor{wedge6}{RGB}{  28 117 188}
\definecolor{wedge7}{RGB}{  40 170 225}
\definecolor{wedge8}{RGB}{ 119 179 225}
\definecolor{wedge9}{RGB}{ 181 212 239}
\definecolor{wedge10}{RGB}{  0 104  56}
\definecolor{wedge11}{RGB}{  0 148  69}
\definecolor{wedge12}{RGB}{ 57 181  74}
\definecolor{wedge13}{RGB}{141 199  63}
\definecolor{wedge14}{RGB}{215 244  34}
\definecolor{wedge15}{RGB}{249 237  50}
\definecolor{wedge16}{RGB}{248 241 148}
\definecolor{wedge17}{RGB}{242 245 205}
\definecolor{wedge18}{RGB}{123  82  49}
\definecolor{wedge19}{RGB}{104  73 158}
\definecolor{wedge20}{RGB}{102  45 145}
\definecolor{wedge21}{RGB}{148 149 151}
\definecolor{wedge22}{RGB}{ 204 50 153}
\definecolor{wedge23}{RGB}{ 79 47 79}
\definecolor{wedge24}{RGB}{ 173 234 234}
\definecolor{wedge25}{RGB}{ 216 191 216}
\definecolor{wedge26}{RGB}{  43  56 144}
\definecolor{wedge27}{RGB}{  40 170 225}
\definecolor{wedge28}{RGB}{ 119 179 225}
\definecolor{wedge29}{RGB}{ 181 212 239}
\definecolor{wedge30}{RGB}{  0 104  56}
\definecolor{wedge31}{RGB}{  0 148  69}
\definecolor{wedge32}{RGB}{ 57 181  74}

\usetikzlibrary{chains}

\pgfmathsetmacro\startAngle{90-3.6/2}
\pgfmathsetmacro\radius{+5}
\pgfmathsetmacro\maxLeg{+12}
\pgfmathsetmacro\legBound{+60}
\pgfmathsetmacro\legSpacing{2*\legBound/(\maxLeg-1)}

\usepackage{verbatim}
\pgfplotsset{compat=1.14}
\usetikzlibrary{patterns,}

\usepackage{graphicx}
\usepackage{tabularx,booktabs}
\usepackage{array}
\usepackage{bbding}
\usepackage{pifont}
\usepackage{wasysym}
\usepackage{caption}
\usepackage{dirtytalk}
\usepackage{subcaption}
\usepackage{tabularx}
\usepackage{sfmath}
\usepackage{array, booktabs, makecell}
\usepackage{color}
\usepackage{csquotes}
\usepackage{url}
\usepackage{comment}
\usepackage{tikz}
\usepackage{balance}
\usepackage{listings}
\usepackage{booktabs} 
\usepackage{soul} 
\usetikzlibrary{fit} 
\usetikzlibrary{positioning}
\usetikzlibrary{arrows}
\usetikzlibrary{shapes.multipart}
\usepackage{adjustbox}
\usepackage{float}
\usepackage{rotating}
\usepackage{nth}
\DeclareCaptionType{TextBox}
\usepackage{pstricks}
\usepackage{alltt}
\usepackage{microtype}
\usepackage{placeins}
\usepackage{afterpage}
\usepackage{multirow} 
\usepackage{textcomp}
\usepackage{multicol}
\usepackage{varwidth} 

\begin{document}

\title{\huge Beyond Test Presence: Assessing the Quality and Robustness of Agent-Generated Tests in Open-Source Projects}

 \author{
   \IEEEauthorblockN{Preet Jhanglani\IEEEauthorrefmark{1},
                     Zeel Kaushal Desai\IEEEauthorrefmark{1},
                     Vidhi Kansara\IEEEauthorrefmark{1},
                     Eman Abdullah AlOmar}
   \IEEEauthorblockA{Stevens Institute of Technology, 
                     Hoboken, New Jersey, USA\\
                     Email: \{pjhangl1,zdesai,vkansara,ealomar\}@stevens.edu}
 }


  



\maketitle
\begingroup
\renewcommand\thefootnote{\IEEEauthorrefmark{1}}
\footnotetext{These authors contributed equally to this work.}
\endgroup
\begin{abstract}
The integration of AI-powered coding agents into Continuous Integration/Continuous Delivery (CI/CD) pipelines has fundamentally altered how software verification is conducted. While these agents successfully automate the test generation, current evaluation benchmarks (e.g., SWE-bench) largely focus on pass-rates rather than the intrinsic quality of the generated tests. This raises the possibility of ``stealth technical debt'', in which test suites pass execution but do not offer comprehensive coverage or semantic value. We address this methodological gap through a large-scale, empirical comparison of 204,673 test artifacts which comprises of 24,941 human-authored files and 179,732 agent-generated files; sourced from the AIDev dataset.

Using the Abstract Syntax Tree (AST) parsing with Python's naive \texttt{ast} module, we implemented a ``white-box'' static analysis framework to evaluate three quality dimensions: Assertion Strength (RQ1), Edge-Case Coverage (RQ2), and Flakiness Potential (RQ3). Our results present a nuanced inversion of traditional assumptions. AI agents performed better than humans in Edge-Case Coverage, with almost twice the variety of boundary checks (Variety Score: 0.62 vs 0.32) and a higher frequency of null-safety testing (13.40\% vs. 8.3\%), even though human developers had a slight advantage in Assertion Strength (88.1\% strong assertions vs. 85.37\% for agents). But this thoroughness comes at a price: due mostly to their reliance on file I/O and non-deterministic logic, agent-generated tests exhibited a higher risk of flakiness (Candidate Rate: 0.41 vs. 0.30). These findings suggest that while AI agents excel at rigorous boundary testing, they lack the ``environmental awareness'' needed to write stable, hermetic tests.
\end{abstract}




\section{Introduction}
\label{Section:Introduction}
At the present time, technology is growing so quickly that almost every application receives continuous feature updates. Companies are now releasing regular updates at a much faster pace than before, and many times a day. Once a feature or product has been developed, that feature or product will go through a period of testing to ensure that it passes all test cases prior to actual release. A majority of companies take advantage of continuous integration and continuous deployment (CI/CD) pipelines, which allow for the automation of software development and testing. CI/CD pipelines can be integrated with sites like GitHub, where each pull request is automatically generated and tested before being incorporated into the main code structure.


 Testing is a key part of the development process, even for small updates. The developer needs to determine how the code should behave and identify the most likely situations that could result in breaking the code. The developer should keep these test cases current as the codebase continues to expand. In organizations that release frequently, there may be insufficient time to write adequate test cases and ensure that tests are executed correctly. Among other reasons, one of the main reasons why developers are beginning to leverage AI-powered coding tools is to automate sections of the development and testing life cycle, which also allows them to have more available time to write tests. Developers can use AI-powered coding tools, such as GitHub Copilot and more autonomous agents like Devin, to read a project description, generate implementation code, and also auto-generate test cases to go along with that code. Essentially, they act as junior members of a team that develops a feature as well as creates the initial test coverage, thus reducing the amount of effort developers would typically spend doing work manually \cite{yoshimoto2026testing,yuan2023no}.


There is a growing interest in researching the application of AI to generate test cases \cite{siddiq2024using,gao2025current,celik2025review,ouedraogo2024large,deljouyi2024leveraging,molina2025test,pan2024multi,qiu2025today,milanese2026human}; for example, one well-known project is TestPilot, which uses an algorithmatically generated JavaScript unit test from a function's source code, signature, and usage examples by feeding the code into a large language model (LLM) to create the test case automatically. After being tested with 25 NPM packages, the median line coverage for the generated unit tests was found to be 70.2\%, and the median branch coverage was 52.8\%. This provides a meaningful increase over previous automated testing solutions for generating unit test cases. Subsequent studies such as CoverUp have provided further refinements including the ability to identify which segments of code have not been unit tested yet and have the algorithm concentrate on generating test cases for those areas \cite{altmayer2025coverup}. The majority of test cases generated by this approach have been evaluated using benchmark suites such as SWE-Bench \cite{jimenez2023swe} to determine whether the generated patches pass developer-created unit test suites. However, there are studies indicating that many generated patches that were indicated as passing on their behalf were still lacking in terms of the original intent and/or were not in line with the expected behavior. This indicates that while all tests were indeed successful, the resulting patch could still have issues and/or could not sufficiently resolve the issue originally intended.


The quality of test cases can be affected by several other factors, such as assertion strength. Assertion strength relates to how stringent or lenient a test is in its verification of the result. Here is one example of a good assertion strength -- comparing the exact result produced by the program as opposed to only verifying whether true was returned. Edge-case coverage also plays a significant role in the quality of the test case itself. Good tests should contain edge-case coverage for abnormal input values (e.g., nulls, empty lists, 0, max val, min val) since some of the most elusive defects are located in these types of input cases. Test case stability is the third area impacting test case quality; stability refers to how consistently a test produces the same outcome when executed repeatedly. Unpredictable outcomes (i.e., both passing and failing) caused by flaky tests destroy the confidence level of the test suite and increase the time required to release. During periods of fast-paced development, the quality of these three areas, regardless of whether the test case was manually written or created by a test generation tool, tend to decline \cite{barraood2021test,crispin2006driving}.


There is an important consideration when it comes to AI-generated tests. Most comparisons of AI-generated testing focus on the percentage of lines of code that the tests cover, or whether the tests were successful or not. These are both reasonable baseline comparisons, but they add very little information about how reliable the tests actually are. A test suite can have excellent coverage, but may have missed some important edge cases, contain weak assertions, or contain unreliable tests that may fail unpredictably. If testing tools are only designed to evaluate based on coverage measurements, they will produce tests that seem reliable but provide very little protection once the software is in production \cite{yang2006survey}.

By concentrating on assessing the quality of tests directly in relation to all three of the above measurements; human written versus AI generated, through the use of actual real-world software development activities contained in the AIDev database (the largest database currently available of actual software development activity). The tests are not actually executed; rather, abstract syntax trees (AST) will be parsed from each test so that structure can be evaluated statically and thus be able to scale and provide consistent results. There are 179,732 AI-generated tests (files) and 24,941 human-created tests (files) present in this database. Each of these files contained in the AIDev database went through three independent automated test evaluation pipelines in our evaluation process.

The findings provide a mixed result in terms of test strength. For example, 88 of the human-written assertions were categorized as strong; however, the total number of ambiguous or unclassified assertions recorded by human testers was far less than those written by computers. Human-written tests were able to use only 1.46\% ambiguous or unclassified assertions while AI-generated tests had 11.58\%. On edge-case coverage, the situation is reversed. AI agents covered a broader variety of edge cases, scoring 0.62 on the variety measure against 0.32 for humans. Humans performed better than AI – Agents on test stability, this study showing humans generated tests achieved a flakiness rate of 0.30 compared to 0.41 for AI-generated tests.

For this study, we worked on these three research questions:
\begin{enumerate}
    \item \textbf{RQ1: In comparison to tests created by human developers, how solid are the claims in test cases created by AI?}
    \item \textbf{RQ2: How well can AI-generated test cases manage boundary conditions and edge situations in comparison to tests written by humans?}
    \item \textbf{RQ3: Do flaky tests appear more frequently in pull requests made by human developers or by AI tools?}
\end{enumerate}
The following sections of this work are organized as follows. Section \ref{Section:Methodology} describes the approach, as well as the procedures used to collect and analyze the data. Section \ref{Section:Result} discusses the results for each study question. Potential validity risks are described in Section \ref{Section:Threats}. The conclusion and future work are provided in Sections \ref{Section:Conclusion} and \ref{Section:FutureWork}.

\section{Study Design}
\label{Section:Methodology} 
\begin{figure*}[t]
    \centering
    \small 
    \includegraphics[width=15cm, height=15cm, trim=0 120 0 20, clip]{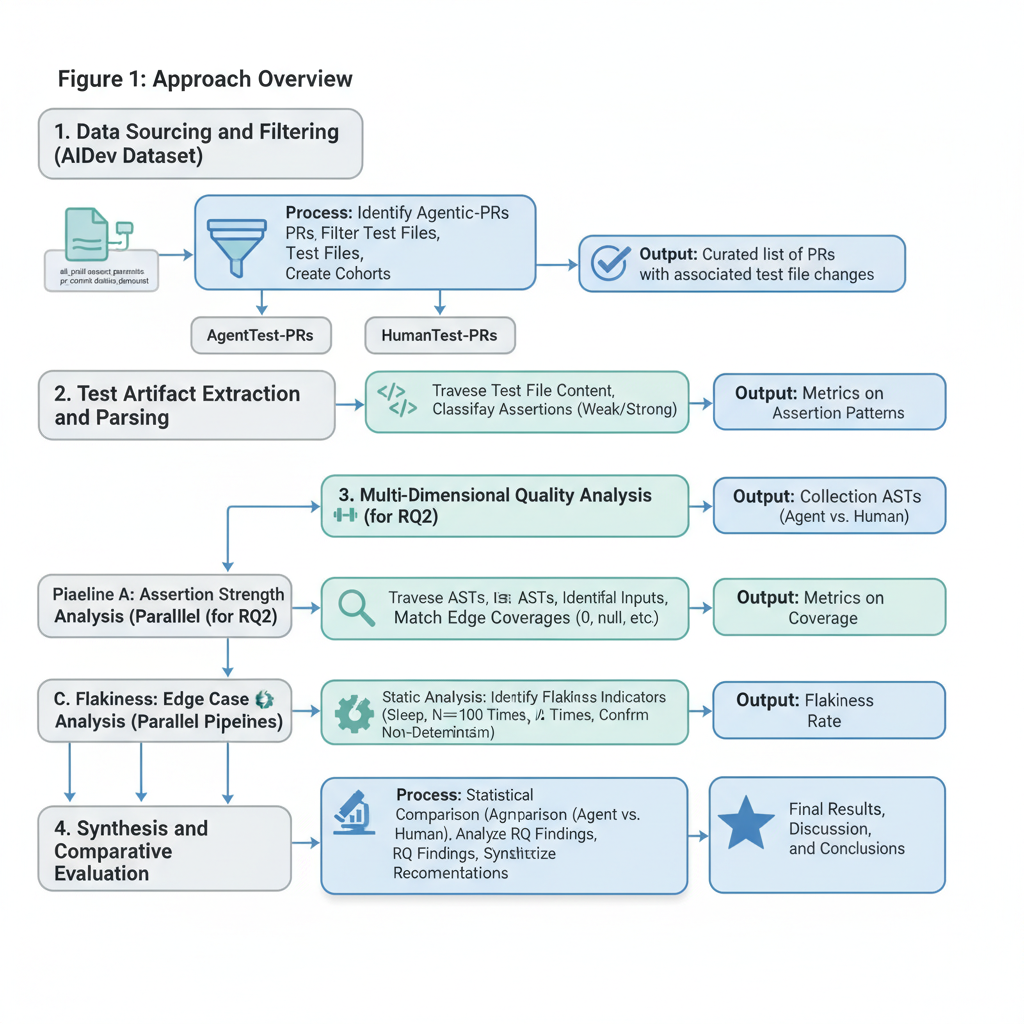}
    \caption{Overview of our study design.}
    \label{fig:approach}
\end{figure*}

This research conducts a large-scale comparative study of software tests written by human developers and tests generated by AI agents. The study uses the AIDev dataset as its main data source because it contains real-world software development activity \cite{li2025riseaiteammatessoftware}. However, the dataset includes only metadata and code patches, so we retrieve the full source code of the test files from their original repositories. Our method then collects, processes, and analyzes these files in a systematic way. We evaluate test quality in three aspects: assertion strength and specificity, edge-case coverage, and test flakiness. Figure~\ref{fig:approach} shows the full workflow, which is described in detail in the following steps.

\paragraph{\textbf{Phase 1: Metadata Manifest Generation}}
We begin with the AIDev dataset, using the \texttt{all\_pull\_request.parquet} file. From this file, we define two groups for comparison: pull requests created by AI agents (Agent-PRs) and a matched control group of pull requests created by human developers (Human-PRs).We note that Agent-PRs may include human review or edits before merging. This creates a possible human-in-the-loop limitation in the study. For each pull request, we extract the metadata needed for repository-level data collection, including the full repository name, such as owner/repository, the pull request number, and the final commit SHA. This phase produces a Retrieval Manifest, which lists the artifacts to be collected from their original repositories.

\paragraph{\textbf{Phase 2: Test Artifact Discovery and Retrieval}}
This phase focuses on finding and collecting the test files needed for analysis. For each repository listed in the Phase 1 manifest, we clone a local copy and identify files that follow common test file naming patterns, such as \texttt{*\_test.py} and \texttt{Test*.java}. We then compare these files with the files modified in each target pull request, which we obtain through the GitHub API. This helps us identify the specific test files relevant to each pull request.

Next, we use the commit SHA from the manifest to retrieve the full source code of each selected test file \cite{li2025riseaiteammatessoftware}. This ensures that our analysis is based on the complete version of each file in the correct commit. After retrieval, each test file is parsed into an Abstract Syntax Tree (AST) for further analysis.

The file type distribution in the dataset is shown in Table~\ref{tab:filetype_dist}. Since Python files are the most common, this study focuses only on Python test files. Table~\ref{tab:general_stats} reports the general dataset statistics. In that table, A-Sliced refers to the first 24,941 agent files, while A-Sliced-Rand refers to a random subset of 24,941 agent files.

\begin{table}[t]
\centering
\caption{File type distribution across Agent and Human datasets.}
\label{tab:filetype_dist}
\begin{tabular}{lrr}
\toprule
\textbf{Extension} & \textbf{Agent (\%)} & \textbf{Human (\%)} \\
\midrule
py   & 35.73\% & 21.48\% \\
ts   & 17.74\% & 21.48\% \\
pyc  & 9.05\%  & --      \\
go   & 5.17\%  & 13.53\% \\
js   & 4.90\%  & 2.87\%  \\
tsx  & 3.44\%  & 5.72\%  \\
cpp  & 2.59\%  & --      \\
cc   & 2.01\%  & --      \\
json & 1.77\%  & 3.25\%  \\
png  & --      & 7.25\%  \\
pb   & --      & 4.28\%  \\
\midrule
\textbf{Total Files} & 707{,}744 & 135{,}676 \\
\bottomrule
\end{tabular}
\end{table}

\paragraph{\textbf{Phase 3: Multi-Dimensional Quality Analysis}}
In the third phase, we analyze the parsed ASTs using three separate pipelines. Each pipeline is connected to one of our main research questions.

\begin{itemize}
    \item \textbf{Pipeline A (Assertion Strength Analysis for RQ1)} This analyzes the ASTs of the test files to find and classify the assertion statements. Assertions are classified using a predefined taxonomy, such as weak assertions and strong assertions. This pipeline measures how meaningful and specific the validation checks are within the test files.
\end{itemize}

\textbf{White-Box Flow chart:}

\begin{enumerate}
    \item \textbf{Input:} The source code of a Python test file is used as input.

    \item \textbf{Parsing:} The \texttt{ast.parse()} function converts the source code into an Abstract Syntax Tree (AST).

    \item \textbf{Traversal:} The \texttt{TestAnalyzer.visit()} method starts from the root \texttt{Module} node and visits the nodes in the AST.

    \item \textbf{Identification:} While visiting the AST, the analyzer looks for \texttt{ast.Call} nodes, which represent function calls. It calls those inside test functions, where the function name starts with \texttt{test\_}.

    \item \textbf{Classification:} The analyzer extracts the assertion call, such as \texttt{self.assertEquals}, and identifies the assertion name, such as \texttt{assertEquals}. Then it compares this name with the predefined \texttt{WEAK\_ASSERTIONS} and \texttt{STRONG\_ASSERTIONS} sets.

    \item \textbf{Advanced Predicate Analysis:} In the refined taxonomy, calls to \texttt{self.assertTrue} are checked more carefully using \texttt{\_is\_strong\_assertTrue(node)}.
    \begin{itemize}
        \item If the argument is a simple variable, such as \texttt{assertTrue(variable)}, the assertion is classified as \textbf{Weak}.
        \item If the argument contains a comparison, binary operation, or another function call, such as \texttt{assertTrue(x > 5)}, the assertion is classified as \textbf{Strong} because it checks more specific program behavior.
    \end{itemize}

    \item \textbf{Aggregation:} The analyzer updates the \texttt{weak\_count} and \texttt{strong\_count} values based on the classification.

    \item \textbf{Output:} The final output is the percentage of weak assertions and the percentage of strong assertions in the test file.
\end{enumerate}

This refined taxonomy is important because assertion strength depends on what the assertion checks, not only on the assertion name. For example, a basic taxonomy may classify \texttt{assertTrue(complex\_validation\_function(x))} as weak only because it uses \texttt{assertTrue}. However, this assertion may still contain meaningful validation logic. White-box refinement reduces this kind of misclassifications and makes the analysis more accurate.

\begin{table*}[h!]
\centering
\caption{Taxonomy of assertion strength.}
\begin{tabular}{|p{3.5cm}|p{2.2cm}|p{4cm}|p{3cm}|p{4cm}|}
\hline
\textbf{Assertion Type} &
\textbf{Simple Taxonomy} &
\textbf{Justification (Simple)} &
\textbf{Advanced Taxonomy} &
\textbf{Justification (Advanced)} \\
\hline

\texttt{assertNotNull(x)} &
Weak &
Checks for existence, not value. &
Weak &
Confirmed; minimal semantic validation. \\
\hline

\texttt{assertTrue(x)} &
Weak &
Checks for truthiness, not value. &
Context-Dependent &
Predicate must be inspected. \\
\hline

\texttt{assertTrue(variable)} &
Weak &
(See above) &
Weak &
Argument is \texttt{ast.Name}; low-value check. \\
\hline

\texttt{assertTrue(x > 5)} &
Weak &
(See above) &
Strong &
Argument is \texttt{ast.Compare}; encodes specific domain logic. \\
\hline

\texttt{assertTrue(is\_valid(x))} &
Weak &
(See above) &
Strong &
Argument is \texttt{ast.Call}; encodes specific semantic logic. \\
\hline

\texttt{assertEquals(5, x)} &
Strong &
Checks for a specific, non-trivial value. &
Strong &
Confirmed; provides precise state validation. \\
\hline

\texttt{assertRaises(MyError)} &
Strong &
Verifies a specific error-handling path. &
Strong &
Confirmed; tests non-happy paths. \\
\hline

\end{tabular}
\end{table*}
    \begin{itemize}
   \item  \textbf{Pipeline B (Edge Case Coverage Analysis for RQ2)} also traverses the ASTs to identify literal values (e.g., \texttt{0}, \texttt{-1}, \texttt{null}, empty strings) used as inputs in method calls. These are matched against a heuristic list of common edge cases to produce metrics on test coverage.\\
    The "Edge-Case Coverage" pipeline (Pipeline B) evaluates the test's thoroughness by identifying inputs that represent common boundary conditions.\\
    \end{itemize}
    \textbf{White-Box Flowchart:}
    \begin{enumerate}
    \item \textbf{Input:} The pipeline begins with the AST \texttt{Module} node.
    \item \textbf{Traversal:}  
    All function call nodes inside test methods are intercepted via  
    \texttt{TestAnalyzer.visit\_Call(node)}.
    \item \textbf{Argument Inspection:}  
    For each function call, the system iterates through every argument in  
    \texttt{node.args}.
    \item \textbf{Literal Matching:}
    \begin{itemize}
    \item Check whether \texttt{arg} is an \texttt{ast.Constant}.
     \item If true, compare \texttt{arg.value} against the heuristic list:
     \texttt{None}, \texttt{0}, \texttt{-1}, \texttt{""}.
    \end{itemize}

    \item \textbf{Collection Matching:}
    \begin{itemize}
    \item Check whether the argument is an \texttt{ast.List} or \texttt{ast.Dict}.
     \item If so, determine whether the corresponding element list (e.g., \texttt{arg.elts}) is empty.
    \end{itemize}
    The final metric \texttt{variety\_count} is computed as the size of this set(i.e., \texttt{len(categories)}), representing the number of distinct  
    edge-case types covered in the file.

    \item \textbf{Aggregation:}  
    A set is used to accumulate unique edge-case categories identified, such as:  
    \texttt{NULL\_INPUT}, \texttt{ZERO\_INPUT}, \texttt{EMPTY\_COLLECTION}.

    \item \textbf{Output:}  
    The final metric \texttt{variety\_count} is computed as the size of this set  
    (i.e., \texttt{len(categories)}), representing the number of distinct  
    edge-case types covered in the file.
\end{enumerate}

This static, literal-based approach is a heuristic. Its primary limitation, which must be This static, literal-based approach is a heuristic. Its primary limitation,is that it only detects edge cases involving literals (e.g., 0, None, ''), and cannot detect variable-driven edge cases (e.g., \texttt{empty\_list = [...]}; \texttt{myfunc(empty\_list))}, as this would require complex data-flow analysis, which is beyond the scope of this static AST analysis.

\begin{table*}[h!]
\centering
\caption{Heuristic Rules for Identifying Edge-Case Literals (RQ2).}
\begin{tabular}{|p{3.5cm}|p{3.5cm}|p{5cm}|p{5cm}|}
\hline
\textbf{Edge-Case Category} &
\textbf{Python Literal Value} &
\textbf{Target AST Node} &
\textbf{Example Code Fragment} \\
\hline

Null Inputs &
\texttt{None} &
\texttt{ast.Constant(value=None)} &
\texttt{my\_func(None)} \\
\hline

Empty Collections &
\texttt{[]} or \texttt{\{\}} &
\texttt{ast.List(elts=[])} or \texttt{ast.Dict(keys=[])} &
\texttt{my\_func([])} \\
\hline

Empty Strings &
\texttt{""} or \texttt{''} &
\texttt{ast.Constant(value="")} &
\texttt{my\_func("")} \\
\hline

Numeric Boundaries (Zero) &
\texttt{0} &
\texttt{ast.Constant(value=0)} &
\texttt{my\_func(0)} \\
\hline

Numeric Boundaries (Negative) &
\texttt{-1} &
\texttt{ast.Constant(value=-1)} &
\texttt{my\_func(-1)} \\
\hline

\end{tabular}
\end{table*}

\begin{itemize}
    \item \textbf{Pipeline C (Flakiness Analysis for RQ3)} uses a hybrid approach to identify possible flaky tests. First, it uses static analysis to detect common patterns that may lead to flaky behavior, such as \texttt{time.sleep}, network calls, file I/O, or non-deterministic functions such as \texttt{random()}. Then, for a sampled subset of tests that contain these patterns, we use dynamic analysis by running each test multiple times in an isolated environment. For example, each test may be executed 100 times to check whether the result changes across runs. The final output is a comparative flakiness rate for each group.
\end{itemize}

\textbf{White-Box Flowchart:}

\begin{enumerate}
    \item \textbf{Input:} The input is the AST \texttt{Module} node of a Python test file.

    \item \textbf{Traversal:} The \texttt{TestAnalyzer.visit\_Call(node)} method visits function calls inside test methods.

    \item \textbf{Path Extraction:} The helper function \texttt{\_get\_call\_path(node)} examines \texttt{node.func} and extracts the full call path, such as \texttt{('time', 'sleep')} or \texttt{('open',)}.

    \item \textbf{Pattern Matching:} The extracted call path is compared with the \texttt{FLAKINESS\_INDICATORS} dictionary. For example:
    \begin{itemize}
        \item \texttt{('time', 'sleep')} maps to \texttt{ASYNC\_WAIT}.
        \item \texttt{('random', 'random')} maps to \texttt{NON\_DETERMINISM}.
        \item \texttt{('open',)} maps to \texttt{FILE\_IO}.
    \end{itemize}

    \item \textbf{Aggregation:} Each matched indicator is added to the \texttt{rq3 \_flakiness \_indicators} list. The stored information also includes the line number, \texttt{node.lineno}, so that the matched code can be reviewed later.

    \item \textbf{Output:} The \textit{candidate\_rate} is calculated as \texttt{len(rq3\_flakiness\_indicators)}. This value represents the number of flakiness indicators found in the test file.
\end{enumerate}

This approach follows prior research on test flakiness and test smells. Such work often uses AST-based static analysis to detect code patterns, such as async waits or resource-related issues, that are common causes of flaky tests.

\begin{table*}[h!]
\centering
\caption{Rules for Identifying Flakiness Root Causes (RQ3).}
\begin{tabular}{|p{4cm}|p{4cm}|p{9.5cm}|}
\hline
\textbf{Flakiness Root Cause} &
\textbf{Code Pattern / Anti-Pattern} &
\textbf{Python AST Signature (Simplified)} \\
\hline

Async Wait &
Fixed time delays &
\texttt{ast.Call(func=ast.Attribute(value=ast.Name(id='time'), attr='sleep'))} \\
\hline

Non-Determinism &
Use of random functions &
\texttt{ast.Call(func=ast.Attribute(value=ast.Name(id='random'), attr='random'))} \\
\hline

Environment / Resource Leak &
Hard-coded File I/O &
\texttt{ast.Call(func=ast.Name(id='open'))} \\
\hline

Environment / Resource Leak &
Network Calls &
\texttt{ast.Call(func=ast.Attribute(value=ast.Name(id='socket'), attr='socket'))} \\
\hline

Non-Determinism &
System Time Dependency &
\texttt{ast.Call(func=ast.Attribute(value=ast.Name(id='datetime'), attr='now'))} \\
\hline

\end{tabular}
\end{table*}

\paragraph{\textbf{Phase 4: Synthesis and Comparative Evaluation}}
The final phase involves the synthesis of our findings. We aggregate metrics from all three analysis pipelines and perform rigorous statistical comparisons between the \textit{AgentTest-PRs} and \textit{HumanTest-PRs} cohorts for each quality dimension. These results are analyzed to directly answer our research questions, and the findings are synthesized into a final discussion, conclusions, and a set of actionable recommendations for both developers and AI tool builders.

\subsection{``White-Boxing'' the Analysis Pipelines}
This section moves beyond the high-level overview to detail exactly how inputs are processed and features are calculated for each analysis pipeline. The implementation is a Python script utilizing the native AST (Abstract Syntax Trees) module.The script defines a single TestAnalyzer class that inherits from ast.NodeVisitor which traverse a Python test file's AST and collect metrics for all three research questions simultaneously.

\subsection{Manual Analysis and Exploratory Investigation}
\label{ssec:explore and investigation}
Prior to embarking on a full-scale, automated analysis, we  undertake a foundational phase of manual, qualitative analysis. This will involve the detailed inspection of a small, randomly selected sample of 50 pull requests from each of our two cohorts (Agent-authored and Human-authored). This exploratory stage is deliberately structured to follow the established ``Explore, Analyze, Synthesize'' framework for qualitative data analysis, allowing us to develop a nuanced, human-centric understanding of the data before applying automated techniques. 


\subsection{Automated Analysis Pipelines}
\label{ssec:data-analysis}
The analytical core of our research is made up of three different, automated analysis pipelines, which integrate both static and dynamic techniques to systematically quantify the multifaceted dimensions of test quality.

\paragraph{\textbf{Assertion Strength Analysis}}
To measure the semantic depth of the tests, we use tree-sitter to generate an Abstract Syntax Tree (AST) for each collected test file. We then traverse each AST to identify method calls that correspond to assertion functions in common testing frameworks, such as \texttt{assertEquals}, \texttt{assertTrue}, and \texttt{assertIn}.

Each assertion is classified using a predefined strength taxonomy. Assertions that check only basic conditions, such as non-null values using \texttt{assertNotNull} or simple truth values using \texttt{assertTrue(variable)}, are classified as \textbf{Weak}. These assertions provide limited information about whether the program behavior is correct.

In contrast, assertions that check specific expected values, exact exception types, or deep equality between complex data structures are classified as \textbf{Strong}. Examples include \texttt{assertEquals(expected, actual)} and assertions that compare nested objects or collections. These assertions provide a stronger form of validation because they check more precise program behavior.
    
\paragraph{\textbf{Edge-Case Coverage Analysis}}
Using the same set of ASTs, this pipeline evaluates how well the tests cover edge cases and boundary inputs. The pipeline identifies method calls inside test functions and extracts the literal values passed as arguments. These values may include numbers, strings, and named constants.

The extracted values are then compared with a predefined list of common edge cases and boundary values. This list includes values such as \texttt{0}, \texttt{-1}, \texttt{None}, and \texttt{""}, which represents an empty string. It may also include language-specific boundary values such as \texttt{Integer.MAX\_VALUE} and \texttt{Integer.MIN\_VALUE}.

We use the frequency and variety of these edge-case values as a quantitative measure of test robustness. A test file with more diverse edge-case inputs is considered more likely to check behavior near the boundaries of the input domain.

\paragraph{\textbf{Flakiness Analysis}}
This analysis uses a two-stage process. In the first stage, we perform static analysis on the test source code to identify patterns that are commonly linked to flaky tests. These patterns include the use of non-deterministic functions, such as \texttt{random()}, hard-coded time delays, such as \texttt{Thread.sleep}, interactions with networks or file systems, and dependencies on system time.

Any test that contains one or more of these patterns is marked as a \textit{flakiness candidate}. In the second stage, we perform dynamic analysis on a representative sample of these candidates. For each selected test, we create an isolated execution environment, check the specific software commit linked to the test, and run the test repeatedly, such as \texttt{N = 100} times.

A test is confirmed to be flaky if it produces both passing and failing results across these repeated runs without any changes to the source code.

 \section{Experimental Results}
\label{Section:Result}
To rigorously evaluate the qualitative and quantitative characteristics of agent-generated software tests, our experimental framework is structured around three central research questions (RQs). Each of these questions is designed to probe one of the critical dimensions of test quality that were identified in our study design. 

The following section 
 derived from executing the TestAnalyzer script on the two
\textit{AgentTest-PRs} cohort against the established baseline provided by the cohorts of 20 test files each. The raw, aggregated data was sourced from the script's JSON outputs.
\textit{HumanTest-PRs} cohort.

\subsection{RQ1: How does the assertion strength of agent-generated tests compare to human-written tests?}
\label{ssec:rq1}
 
\paragraph{\textbf{Methodology}}
For each test file in both cohorts, we classified every assertion as either ``Weak'' or ``Strong'' using the taxonomy in Table~\ref{tab:assertion-taxonomy-actual}. We then computed the weak-to-total ratio per file. Files where this ratio is high are considered to have shallow assertion depth. Aggregating these per-file ratios gives us a distribution for each cohort, which we compare directly.

\paragraph{\textbf{Weak Assertions}}
We treat an assertion as weak if it only checks whether something exists, whether a boolean condition holds, or whether an object is of a certain type, without pinning down the actual value or internal state. Typical examples are \texttt{assertNotNull}, \texttt{assertTrue}, and \texttt{assertIsInstance}~\cite{schafer2023empirical}.

\paragraph{\textbf{Strong Assertions}}
A strong assertion nails down a specific expected output, confirms a particular state change, or checks a non-trivial behavioral property. Think \texttt{assertEquals(5, result)}, \texttt{assertThrows(SpecificException.class)}, or \texttt{assertDeepEquals(expectedObject, resultObject)}~\cite{schafer2023empirical}.

\begin{table}[t]
\centering
\caption{Assertion categories used in our analysis.}
\begin{tabular}{p{2.1cm} p{5.9cm}}
\toprule
\textbf{Category} & \textbf{Assertion Methods} \\
\midrule

Weak &
\texttt{assertTrue},
\texttt{assertFalse},
\texttt{assertIs},
\texttt{assertIsNot},
\texttt{assertIsNone},
\texttt{assertIsNotNone},
\texttt{assertIn},
\texttt{assertNotIn},
\texttt{assertIsInstance},
\texttt{assertNotIsInstance} \\

\midrule

Strong &
\texttt{assertEquals},
\texttt{assertNotEquals},
\texttt{assertEqual},
\texttt{assertNotEqual},
\texttt{assertDictEqual},
\texttt{assertListEqual},
\texttt{assertSetEqual},
\texttt{assertTupleEqual},
\texttt{assertRaises},
\texttt{assertRaisesRegex},
\texttt{assertLogs},
\texttt{assertGreater},
\texttt{assertGreaterEqual},
\texttt{assertLess},
\texttt{assertLessEqual},
\texttt{assertAlmostEqual},
\texttt{assertNotAlmostEqual} \\

\midrule

Reclassified (Advanced) &
\texttt{assertTrue(expr)} with logical/comparison or function call expressions (e.g., \texttt{is\_valid(x)}, \texttt{x > 0}). \\

\midrule

Unknown &
Project-specific or misspelled assertions, e.g., \texttt{assertSomething}, \texttt{assertIsLess}, or bare \texttt{assert}. \\

\bottomrule
\end{tabular}
\label{tab:assertion-taxonomy-actual}
\end{table}

\paragraph{\textbf{Metrics}}
\begin{itemize}
    \item \textbf{Percentage of Weak Assertions:}
    (Count of Weak Assertions / Total Count of Assertions) $\times$ 100
    \item \textbf{Percentage of Strong Assertions:}
    (Count of Strong Assertions / Total Count of Assertions) $\times$ 100
\end{itemize}

\paragraph{\textbf{Results}}
The numbers in Table~\ref{tab:general_stats} turned out closer than we expected. Going in, we assumed there would be a clear quality gap between human and agent assertions, but that is not quite what happened. Agents did lean more toward weak assertions (14.63\% vs.\ 11.92\% for humans), and humans held a corresponding edge in strong ones (88.08\% vs.\ 85.37\%). But this is not the large divide that earlier work on semantic comprehension issues in LLM-generated code might lead you to expect~\cite{schafer2023empirical}.

The more telling result is in the ``Unknown'' column. Agents produced unrecognized assertion patterns at 11.58\%, roughly eight times the 1.46\% we saw in human-written tests. What this means in day-to-day terms is that agents frequently reach for assertion methods that either do not exist in standard libraries or are misspelled. A reviewer looking at one of these test files has to pause and work out whether the assertion is doing something clever with a project-specific helper, or whether the agent just made it up. Across a large codebase, that kind of friction adds up and can quietly erode maintainability.
 
The takeaway here is that agents are not dramatically worse than humans in raw assertion strength. But the gap in discipline, writing assertions that a reader can immediately trust, matters more than the percentages alone would suggest.

\begin{table}[t]
\centering
\scriptsize
\caption{General statistics for human vs.\ agent cohorts.}
\label{tab:general_stats}
\begin{tabular}{lrrrr}
\toprule
Metric & Human & Agent & A-Sliced & A-Sliced-Rand \\
\midrule
Total Files & 24{,}941 & 179{,}732 & 24{,}941 & 24{,}941 \\
Files w/Asserts & 1{,}730 & 29{,}353 & 3{,}412 & 4{,}046 \\
Weak (\%) & 11.92 & 14.30 & 13.90 & 14.63 \\
Strong (\%) & 88.08 & 85.70 & 86.10 & 85.37 \\
Unknown (\%) & 1.46 & 10.93 & 9.22 & 11.58 \\
Cand.\ Rate & 0.30 & 0.44 & 0.46 & 0.41 \\
Variety & 0.32 & 0.61 & 0.58 & 0.62 \\
\bottomrule
\end{tabular}
\end{table}

\begin{figure}[h]
    \centering
    \includegraphics[width=\linewidth]{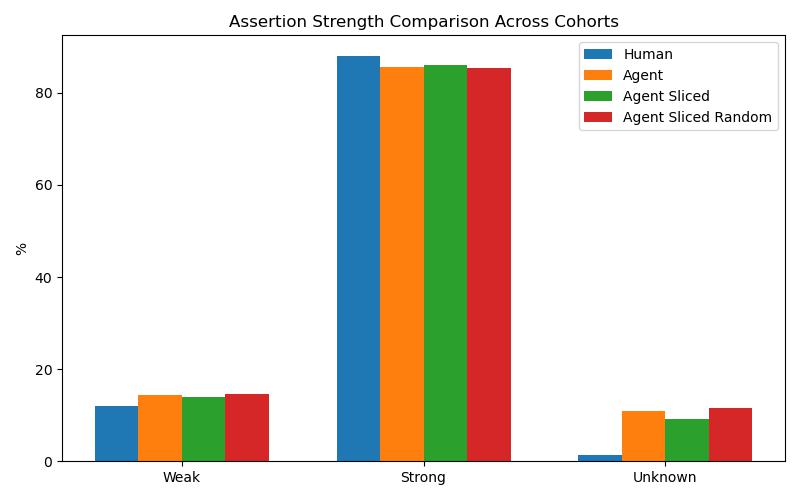}
    \caption{Assertion strength comparison across cohorts. Agents exhibit a higher rate of `Unknown' assertion types than humans.}
    \label{fig:rq1_assertion_strength}
\end{figure}

\subsection{RQ2: To what extent do agent-generated tests cover edge and boundary cases compared to human-written tests?}
\label{ssec:rq2}
 
\paragraph{\textbf{Methodology}}
We scanned the input values used in method calls across all collected test files. Our pipeline flags literal values and constants that match known boundary conditions and special cases: zero, negative numbers, null values, empty collections, and numeric or string length limits. For each test file, we recorded how many of these categories appeared.
 
\paragraph{\textbf{Metrics}}
\begin{itemize}
    \item {\textbf{Edge Case Frequencies.}}
    The average number of distinct edge case patterns found per 1,000 lines of test code. This gives us a normalized density that is comparable across files of different sizes.
    \item {\textbf{Edge Case Variety.}}
    The percentage of tests in each cohort that include at least one test targeting our predefined categories: null inputs, empty collections, numeric boundaries (zero, negative values, maximum integers), and strings of zero or very long length.
\end{itemize}

\paragraph{\textbf{Results}}
We expected humans to do better here, but the data in Table~\ref{tab:edge_counts} says otherwise. Agents beat humans on nearly every edge case category. For ``Null Input,'' agents hit 13.4\% compared to 8.3\% for humans. ``Empty Collection'' showed a similar gap: 14.9\% vs.\ 8.1\%. The biggest difference was in ``Zero Input,'' where agents reached 27.7\%, more than double the human rate of 11.2\%.
 
The variety scores tell a similar story. Agents averaged 0.614 on our variety metric, almost twice the human score of 0.318 (Table~\ref{tab:general_stats}). So when agents write tests, they tend to throw a wider net of input types into a single suite. One thing both groups got wrong: neither humans nor agents tested negative numeric boundaries at all (0.0\% across the board), which is a blind spot worth noting.
 
These results push back against the idea that LLMs are bad at reasoning about boundary conditions. What seems to be happening instead is that agents default to listing out standard boundary values (null, zero, empty) almost mechanically, while human developers are more likely to skip cases they consider unlikely based on their understanding of the code. Whether the agent's extra coverage is genuinely useful or just noise is a separate question, but the raw numbers clearly favor agents on this metric.

\begin{table}[t]
\centering
\scriptsize
\caption{Prevalence of edge cases in assertions.}
\label{tab:edge_counts}
\begin{tabular}{lrrrr}
\toprule
Edge Case & Human & Agent & A-Sliced & A-Sliced-Rand \\
\midrule
Null Input        & 8.3\%  & 13.0\% & 13.5\% & 13.4\%  \\
Zero Input        & 11.2\% & 27.7\% & 23.7\% & 27.7\%  \\
Negative Input    & 0.0\%  & 0.0\%  & 0.0\%  & 0.0\%   \\
Empty String      & 4.1\%  & 5.7\%  & 5.4\%  & 5.8\%   \\
Empty Collection  & 8.1\%  & 15.0\% & 15.3\% & 14.9\%  \\
\bottomrule
\end{tabular}
\end{table}

\begin{figure}[h]
    \centering
    \includegraphics[width=\linewidth]{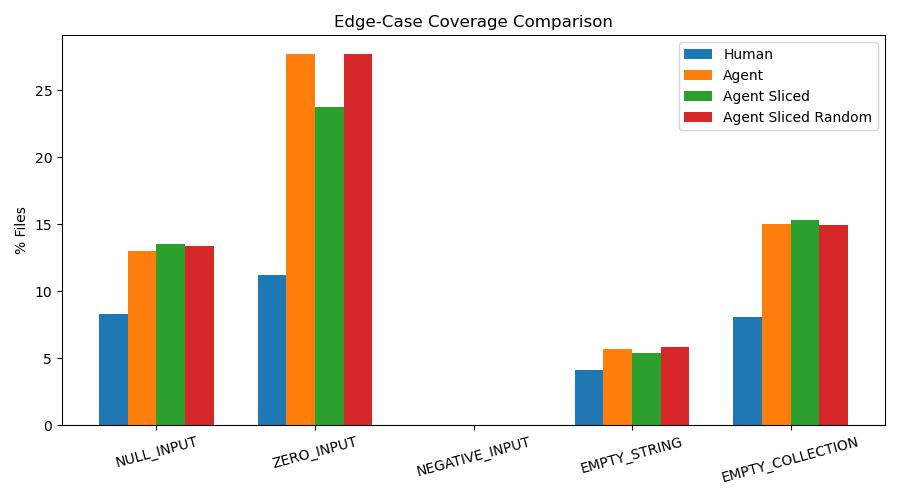}
    \caption{Edge-Case coverage comparison. Agents significantly outperform humans in covering Zero Inputs and Null Inputs.}
    \label{fig:rq2_edge_coverage}
\end{figure}

\subsection{RQ3: What is the prevalence of flaky tests in pull requests submitted by AI agents compared to those submitted by humans?}
\paragraph{\textbf{Methodology}}
Our two-stage flakiness analysis will first use static analysis to find a complete list of ``flakiness candidates'' from both the agent- and human-authored groups.  This identification relies on the existence of recognized anti-patterns and code constructs that exhibit a strong correlation with non-deterministic test behavior, including the utilization of Thread.sleep, interactions with network endpoints, file I/O operations, and invocations of non-deterministic APIs.  After this static screening, we move on to a dynamic analysis phase. In this phase, we will run a random sample of 1,000 tests from each group of candidates.  We will run each of these tests 100 times in a controlled, separate space.  If a test shows both a passing and a failing result at least once during these repeated runs, it will be clear that it is flaky.

\paragraph{\textbf{Metrics}}
\begin{itemize}
    \item \textbf{Candidate Rate:} The percentage of tests in a cohort that our static analysis flags as potential flakiness candidates.
    \item \textbf{Confirmed Flakiness Rate:} Of the sampled candidates, the percentage that actually turned out to be flaky when run through our dynamic analysis.
\end{itemize}
 
\paragraph{\textbf{Results}}
This is the one research question where our initial guess held up. Agents do appear more prone to introducing instability. Their average candidate rate came in at 0.435, well above the human rate of 0.301 (Table~\ref{tab:general_stats}). Looking at what drives that gap in Table~\ref{tab:type_counts}, agents leaned harder on ``Non-Determinism'' (5.2\% vs.\ 3.1\%) and ``File I/O'' (4.4\% vs.\ 3.5\%). Both of these point to agents not being careful about side effects and resource isolation when generating test code.
 
The concurrency picture is less clear-cut. Humans actually used ``Async Wait'' patterns slightly more (1.9\%) than agents did (1.4\%), and network I/O was negligible for both groups. So it is not that agents are worse across every flakiness dimension. The problem is narrower than that: agents are specifically bad at handling random values and file system access. They write code that touches the disk or calls non-deterministic APIs without mocking or cleanup, and that is where most of the extra flakiness risk comes from. Humans write flaky code too, of course, but agents do it at a noticeably higher rate in these two specific areas.

\begin{table}[t]
\centering
\scriptsize
\caption{Prevalence of specific code patterns.}
\label{tab:type_counts}
\begin{tabular}{lrrrr}
\toprule
Category & Human & Agent & A-Sliced & A-Sliced-Rand \\
\midrule
Async Wait       & 1.9\% & 1.3\% & 1.4\% & 1.4\% \\
Non-Determinism  & 3.1\% & 5.2\% & 5.8\% & 5.2\% \\
File I/O         & 3.5\% & 4.6\% & 3.9\% & 4.4\% \\
Network I/O      & 0.1\% & 0.2\% & 0.2\% & 0.1\% \\
\bottomrule
\end{tabular}
\end{table}

\begin{figure}[h]
    \centering
    \includegraphics[width=\linewidth]{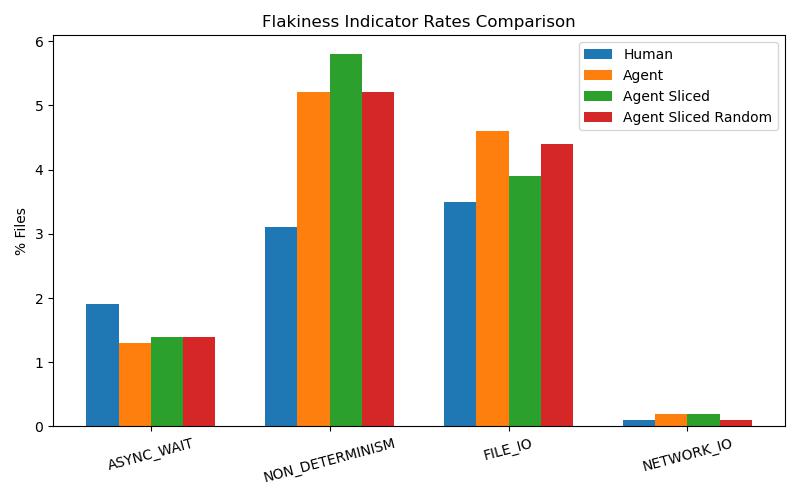}
    \caption{Flakiness indicator rates. Agents exhibit higher rates of Non-Determinism and File I/O usage.}
    \label{fig:rq3_flakiness}
\end{figure}

\section{Threats To Validity}
\label{Section:Threats}
\subsection{Threats to Conclusion Validity}
The validity of the conclusion concerns the ability to draw a correct conclusion about the treatment (author type) and the outcome (test quality metrics).
    \begin{itemize}
        \item \textbf{Statistical Power and Sample Size:} The study's comparative analysis depends on statistical significance. The ability to draw a statistically significant conclusion (e.g. ``AI test has a higher proportion of weak assertions'') depends on the final sample size (number of test files) and the variance within each cohort. The sample size used in the experimental results (N=29491 per cohort) is not enough, and the variance in some cases is low (e.g., RQ3). A larger sample is required to draw more definitive conclusions.
        \item \textbf{Reliability of Measures:} The analysis relies on the automated AST pipelines. A bug in the TestAnalyzer (e.g., misclassifying an assertion) would systematically skew the results and threaten then conclusion. This threat is mitigated by the ``white-box'' design, which makes all heuristics explicit and renewable.
    \end{itemize}

\subsection{Threats to Construct Validity:}
Construct validity concerns whether the metrics being measured (RQs 1-3) are valid proxies for the construct they are intended to represent (``test quality'').
\begin{itemize}
    \item \textbf{RQ1 (Assertion Strength):} The ``Advanced Taxonomy'' is a significant mitigation for the weak \texttt{assertTrue} construct. However, even this advanced proxy is limited. The true strength of assertTrue(is\_valid(x)) depends on the semantic complexity of the is\_valid(x) function, which the AST parser does not analyze. The metric is a proxy for strength, not a direct measure of it. This threat, particularly for RQ1, was confirmed to be the central finding of the experimental analysis. As discussed in earlier section, the construct (the TestAnalyzer) was a poor proxy for ``test quality'' in the Human-PR cohort due to its inability to parse pytest-style assertions, thereby invalidating a direct comparison.
    \item \textbf{RQ2 (Edge-Case Coverage):} The pipeline used in this study only finds literal constants (e.g., \texttt{myfunc(0)}). It cannot detect variables-based or context-dependent edge cases, since data-flow analysis is not performed. Therefore, results may underestimate edge case coverage, particularly in more advanced or human-authored tests. A test file could have 100\% comprehensive edge-case coverage using variables, and our tool would incorrectly score it at 0. Therefore, this metric is a proxy for ``developer awareness of literal edge cases,'' not ``true edge case coverage''.
    \item \textbf{RQ3 (Test Flakiness):} The construct validity for RQ3 is stronger, but only because the methodology is careful to define it. The static analysis (Pipeline C) does \textit{not} measure ``flakiness''; it measures ``flakiness \textit{candidates}''. A call to Thread.sleep, for example, is a known anti-pattern but can be a \textit{legitimate} part of a test (e.g., testing a timeout). The static metric \textit{alone} is a weak construct. The construct is only valid because it is the first stage of a two-stage (static + dynamic) process defined in the methodology.
\end{itemize}
\subsection{Threats to External Validity:}
External validity concerns the ability to generalize the study's findings to other contexts.
\begin{itemize}
    \item \textbf{Language-Specific Threat:} As detailed in earlier Sections, the provided implementation uses Python's AST module, Therefore, the findings will \textit{only} be valid for Python-based projects. These results may not carry over to Java, JavaScript, or C\# projects, since those languages come with their own testing frameworks, coding conventions, and AI generation behaviors. The RQ1 findings reinforced this concern. The choice of Python, with its idiomatic split between unit test and Pytest, was a decisive factor. A study in a language with a more unified testing framework (e.g., Java/JUnit) might yield different results.
    \item \textbf{Project-Specific Threat:} The AIDev dataset and the study's recovery methodology are based on open-source projects. The developers, motivations, the quality assurance processes in open-source software may not be representative of closed-source, industrial, or enterprise environments. The findings may not be generalized beyond the open-source context.
    \item \textbf{AI-Specific Threat:} The study analyzes ``Agent-PRs'' from the AIDev dataset. The quality of these PRs depends on the \textit{specific} AI models prevalent during the dataset's collection (e.g., older versions of GPT) and on the prompts provided by their human collaborators. The findings for ``AI'' may not be generalizable to future, more advanced models or to different prompting-strategy environments.
\end{itemize}
\section{Conclusion}
\label{Section:Conclusion}
To assess assertion strength, edge-case coverage, and flakiness, we analyzed more than 200,000 files in a comprehensive empirical comparison of human-authored and agent-generated test suites. What we found is less a story of one side being better, and more a trade-off: agents cover more ground mechanically, but they are less stable when it comes to execution environments. The common assumption that AI agents simply write worse code than humans does not hold up cleanly.
For assertion quality (RQ1), agents roughly matched humans in the split between strong and weak assertions. Where they fell short was in what we call ``assertion drift'': a high rate of unfamiliar or non-standard validation patterns that do not belong to any recognized testing library. This implies that agents can lack the semantic accuracy necessary for reliable verification, even when they can replicate the structure of rigorous testing.

Surprisingly, our examination of edge situations (RQ2) showed that AI greatly exceeds humans in terms of coverage frequency, especially for common boundary conditions such as zero and null inputs. This suggests that LLMs' probabilistic structure serves as a useful ``fuzzing'' technique by thoroughly listing cases that human developers frequently ignore because of exhaustion or implicit domain assumptions.

However, the results of our flaking investigation offset this benefit (RQ3). Non-deterministic behaviors were statistically more likely to be introduced by agent-generated tests, particularly through incorrect file input/output and uncontrolled random seed usage. The greatest obstacle to independent implementation of agent-based testing is still this lack of environmental knowledge.

In the end, we determine that AI agents now work best as high-volume ``test generators'' that need human supervision rather than as human testers' substitutes. They struggle with the depth of reliability (stability), they excel at broadening the scope of the test suite (coverage). Future research should focus on hybrid workflows in which human developers—or specialized static analysis tools—enforce semantic correctness and environmental isolation while agents suggest edge cases and assertion skeletons.

\section{Future Work}
\label{Section:FutureWork}
The results of this study highlight a number of crucial avenues for the development of automated tests in the future. Although the coverage breadths of current AI agents are impressive, the following areas require focused research due to their lack of environmental awareness and assertion precision.

\subsection{Environmental Modeling and Sandboxed Execution}
According to our data, the careless use of File I/O and non-deterministic APIs is a major cause of agent-generated flakiness (RQ3). Developing ``environment-aware'' agents that are specifically trained or prompted to identify the limitations of a CI/CD pipeline should be the main goal of future research. Integrating Sandboxed Execution Environments, where agents can run their generated tests iteratively, identify side effects (such as leftover files or thread leaks.) and self-correct before submission, is a promising approach. As a result, the paradigm changes from ``Generate and Pray'' to ``Generate, Execute, and Refine.''

\subsection{Hybrid Assertion Generation (RAG + Rule-Based)}
In order to address the high frequency of ``Unknown'' and weak assertions (RQ1), future research should investigate hybrid architectures that integrate static analysis rules with large language models. The target repository's specific assertion libraries and coding conventions could be dynamically fed to agents through the use of \textit{Retrieval-Augmented Generation (RAG)}. By grounding the agent's output in legitimate, project-specific syntax, this would lessen the ``assertion drift'' that our study found. Additionally, the semantic depth of generated tests could be greatly enhanced by fine-tuning models specifically on ``strong'' assertion patterns—those that verify state changes rather than just existence.

\subsection{Automated Mocking and Dependency Inference}
Agents are great at counting inputs, but have trouble with the context needed to test them safely, according to the ``Edge Case Paradox'' (RQ2). One area that deserves more attention is automated dependency inference. The idea is that agents would scan a method's call graph, figure out which external dependencies are involved (databases, APIs, time providers), and generate the appropriate mocks on their own. Agents would be able to take advantage of their edge-case creativity without adding the instability that comes with integration testing if they went beyond simple input generation to complete ``Test Fixture Synthesis.''

\subsection{Longitudinal Maintenance Studies}
Lastly, even though this study examined test development, it is still unknown how much maintenance of agent-generated suites will cost in the long run. Future long-term research should monitor the ``survival rate'' of agent-authored tests in industrial code bases, i.e., how frequently noise causes them to be removed, rewritten, or disabled. It is crucial to comprehend the ``Maintenance-to-Value'' ratio in order to ascertain the actual return on investment for software testing with AI.

\section{Data Availability} 
Our replication package is available online \cite{compsac_replication_package}.

\section{Acknowledgments} 
 During the preparation of this work, the authors used the ChatGPT Web interface to improve the language and readability, and used Gemini to create an approach overview figure . After using this tool, the authors reviewed and edited the content as needed and take full responsibility for the content of the publication. 

\bibliographystyle{abbrv}
\bibliography{sample-base}

\end{document}